\documentclass[conference]{IEEEtran}
\IEEEoverridecommandlockouts
\usepackage{cite}
\usepackage{amsmath,amssymb,amsfonts}
\usepackage{algorithmic}
\usepackage{graphicx}
\usepackage{textcomp}
\usepackage{xcolor}
\usepackage{svg}
\usepackage{listings}
\usepackage{subcaption}
\usepackage{graphicx}
\usepackage{fancyvrb}
\usepackage{multirow}
\def\BibTeX{{\rm B\kern-.05em{\sc i\kern-.025em b}\kern-.08em
    T\kern-.1667em\lower.7ex\hbox{E}\kern-.125emX}}
    
\begin{document}

\title{Hybrid Cloud and HPC Approach to High-Performance Dataframes\\

}

\author{
\IEEEauthorblockN{Kaiying Shan\IEEEauthorrefmark{1}, 
Niranda Perera\IEEEauthorrefmark{2}, 
Damitha Lenadora\IEEEauthorrefmark{3}, 
Tianle Zhong\IEEEauthorrefmark{1},
Arup Kumar Sarker\IEEEauthorrefmark{1},
Supun Kamburugamuve\IEEEauthorrefmark{4},\\
Thejaka Amila Kanewela\IEEEauthorrefmark{4}, 
Chathura Widanage\IEEEauthorrefmark{4}, 
Geoffrey Fox\IEEEauthorrefmark{1}\IEEEauthorrefmark{5}
}
\IEEEauthorblockA{
\IEEEauthorrefmark{1}University of Virginia, Charlottesville, VA 22904, USA
\{ks5qug, fad3ew, djy8hg\}@virginia.edu}
\IEEEauthorblockA{\IEEEauthorrefmark{2}Luddy School of Informatics, Computing, and Engineering, Indiana University,\\
Bloomington, IN 47408, USA\\
dnperera@iu.edu}
\IEEEauthorblockA{\IEEEauthorrefmark{3}University of Illinois Urbana-Champaign, Urbana, IL 61801, USA\\
damitha2@illinois.edu}
\IEEEauthorblockA{\IEEEauthorrefmark{4}Indiana University Alumni, Bloomington, IN 47405, USA\\
supun@apache.org, \{thejaka.amila, chathurawidanage\}@gmail.com}
\IEEEauthorblockA{\IEEEauthorrefmark{5}Biocomplexity Institute and Initiative, University of Virginia, Charlottesville, VA\\
22911, USA\\
vxj6mb@virginia.edu}
}

\maketitle

 {\let\thefootnote\relax\footnote { {Kaiying Shan and Niranda Perera contributed equally to this work. \\978-1-6654-8045-1/22/\$31.00 ©2022 IEEE}}}
\begin{abstract}
Data pre-processing is a fundamental component in any data-driven application. With the increasing complexity of data processing operations and volume of data, Cylon, a distributed dataframe system, is developed to facilitate data processing both as a standalone application and as a library, especially for Python applications. While Cylon shows promising performance results, we experienced difficulties trying to integrate with frameworks incompatible with the traditional Message Passing Interface (MPI). While MPI implementations encompass scalable and efficient communication routines, their process launching mechanisms work well with mainstream HPC systems but are incompatible with some environments that adopt their own resource management systems. In this work, we alleviated this issue by directly integrating the Unified Communication X (UCX) framework, which supports a variety of classic HPC and non-HPC process-bootstrapping mechanisms as our communication framework. While we experimented with our methodology on Cylon, the same technique can be used to bring MPI communication to other applications that do not employ MPI's built-in process management approach.

\end{abstract}

\begin{IEEEkeywords}
Distributed Dataframe, High-Performance Computing, Cloud Computing, UCX
\end{IEEEkeywords}


\section{Introduction}
Data engineering pipelines are used pervasively in both industry and academia, and dataframe serves as a key component in the practice. 
As the scale of data increases, distributed runtime libraries, such as Dask \cite{rocklin2015dask} and Ray \cite{ray}, emerged and were widely adopted, and they significantly alleviated the complexity of working in large clusters. 
On top of these distributed runtime libraries, distributed dataframe (DDF) solutions like Dask DDF, Ray Datasets, and Modin are proposed. 
Cylon\cite{widanage2020high} is also a distributed dataframe system but outperforms its competitors in many scenarios, especially for High-Performance Computing (HPC).

Using MPI as its communication framework, Cylon can achieve a promising scaling ability and run on various hardware. 
MPI implementations come with efficient and scalable process launching and management mechanisms, such as mpirun\cite{mpirun}.
However, while the mechanisms are efficient and scalable, they are incompatible with some distributed runtime libraries, including Dask and Ray, as such libraries employ their built-in resource management mechanisms.
While MPI implementations lean towards the HPC ecosystem, the distributed runtime libraries and frameworks also embrace the distributed and cloud computing technology stack, such as Kubernetes\cite{kubernetes}.
This paper presents an effort to make Cylon able to be incorporated into the distributed runtime libraries by integrating UCX \cite{shamis2015ucx} and its complementary collective operations application programming interface (API) and library, Unified Collective Communication (UCC)\cite{ucc}. 






MPI implementations usually come with process bootstrapping functionalities that handle process launching and management. 
These functionalities help set up the information necessary for communication: world size, rank, and other processes' contact addresses. 
In contrast, UCX doesn't enforce any process management and serves solely as a network communication library. 
This characteristic aligns with Cylon's goal as a library for distributed computing applications, as those applications usually have their own mechanism for configuration and management.
With both traditional MPI and UCX as options for the network communication library, Cylon is able to serve both in the context of HPC and cloud computing.



This paper will discuss how we integrated UCX into Cylon's communication and distributed execution model.
Furthermore, as UCX does not have its own bootstrapping mechanism, we also developed a mechanism to initiate UCX/UCC communication without incompatible dependencies. 
With this integration, Cylon can be integrated with Dask and Ray, unlocking many potential improvements in the data engineering pipeline. 
We also managed to achieve  performance at least as good as when using traditional MPI.

\section{Motivation}
\subsection{Cylon Overview}
The work presented in this paper is an extension of Cylon. Cylon is a high-performance distributed-memory framework parallelized using the Bulk Synchronous Parallel (BSP) pattern \cite{Tiskin2011}. It can load and process heterogeneously structured data efficiently. Cylon was developed with the ideology that high-performance data processing should be available in a wide range of scenarios that involves a large amount of data, including data engineering, AI/ML applications, distributed databases, and so on. It is intentionally developed to be available as both a library that provides functions to load, extract, and transform data efficiently and as a framework that can run in a standalone fashion.
\subsection{Cylon Architecture}

\begin{figure}[btp]
\centering
\includegraphics[width=\linewidth]{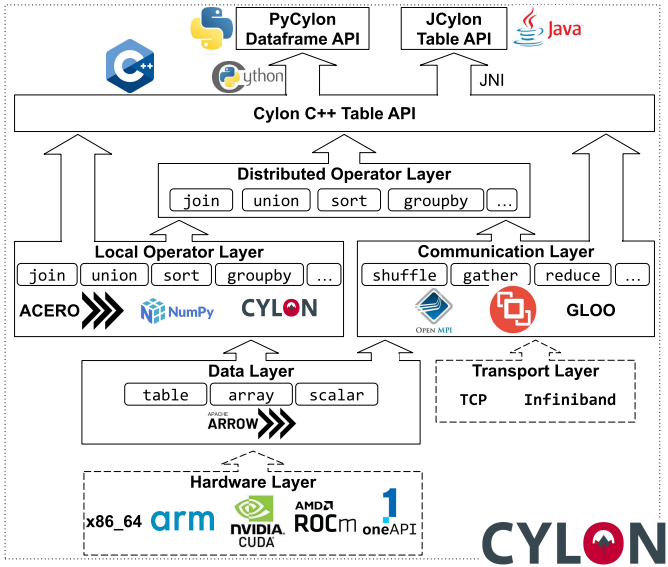}
\caption{Cylon Architecture}
\label{fig:cylon-arch}
\end{figure}

A Cylon dataframe represents one dataframe partitioned into a group of dataframes that may exist on different processes, machines, or clusters. The operators of such dataframes apply on each partition simultaneously with a Single Program Multiple Data (SPMD) pattern and use the collective operation to perform interaction between processes. Figure \ref{fig:cylon-arch} presents an overview of Cylon's multilayered architecture.
Cylon consists of several data abstractions: dataframe/table, column, and scalar, and two levels of procedural abstractions: dataframe operators and communication operators. In this context, we refer to a column as an array of scalars of the same data type, and we refer to that data type as the data type of the column. We use the term \textbf{dataframe} interchangeably with \textbf{table}, as we refer to both of them as the data structure that consists of multiple columns of possibly different data types.

\begin{figure}[btp]
\centering
\includegraphics[width=\linewidth]{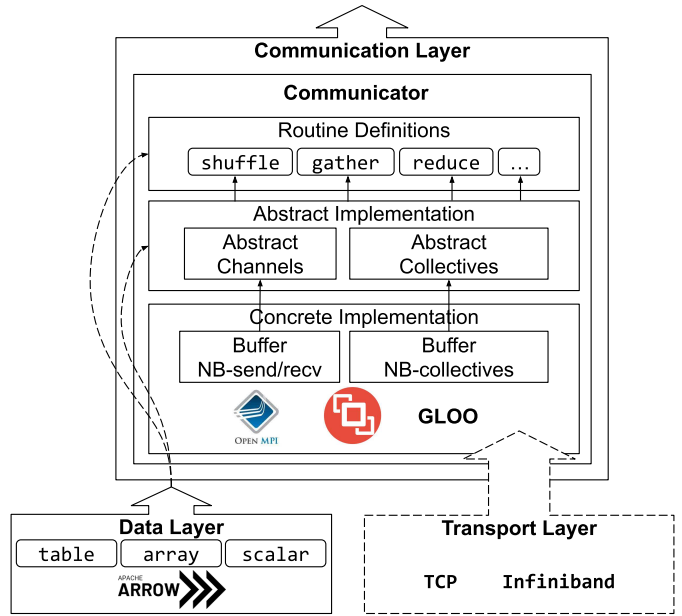}
\caption{Cylon Communication Architecture}
\label{fig:cylon-comm-arch}
\end{figure}

\subsubsection{Cylon Communicator}
The Communicator interface provides access to the communication operators and hides the complexities of communication frameworks. It abstracts out point-to-point communication and collective communication routines.
These collective communication operators, such as \texttt{AllToAll}, \texttt{AllGather}, \texttt{AllReduce}, \texttt{Gather}, \texttt{Broadcast}, etc., are selectively implemented for the data structures (dataframe, column, and scalar). Unlike homogeneous arrays and tensors, dataframes consist of heterogeneous structures and involve variable length buffers. Therefore, it requires a combination of communications to perform each communication operator rather than directly calling the communication framework APIs.

The communicators lay on top of the network communication framework layer, which actually handles the exchange of data over transport layer protocols such as TCP and Infiniband. In the original Cylon implementation, the network communication layer consists solely of MPI. In recent versions, we added integration of other communication frameworks: Gloo and UCX. Figure \ref{fig:cylon-comm-arch} illustrates the layered structure and relationships of dataframe operators (Table API), communication operators, and communication frameworks.

Cylon also supports loading from multiple file formats, including CSV, JSON, and Apache Parquet. When creating a Cylon table in a distributed context, data is automatically loaded and partitioned towards each worker, preparing for further parallel processing.

\subsection{Challenges}
As Cylon is able to serve as a library, we envision it to be integrated with other data engineering or machine learning frameworks to help process data efficiently. For example, Ray is a distributed computing framework consisting of a distributed runtime and a machine learning library, able to scale and accelerate ML/AI applications. Another example is Dask, a distributed computing framework with task scheduling to scale Python applications. Both frameworks provide solutions to scale big data applications in Python effortlessly. Following Cylon's ideology of making high-performance data processing available to every big-data scenario, we want to enable it to load and process data efficiently for applications using these frameworks. However, as Cylon uses MPI as its communication interface, we spot the incompatibility issue when trying to integrate Cylon with these frameworks.

We find UCX with UCC as an appropriate substitute for MPI. As an interface, MPI can build on many communication frameworks, including UCX. However, UCX can also operate as a communication framework itself. Like MPI, UCX and UCC are efficient and able to run on a variety of hardware and communication protocols. As UCX and UCC do not have their own bootstrapping mechanism, MPI is the \textit{de facto} mechanism for them. To break this dependency, we implemented a more flexible initialization process. We modularized the bootstrapping process and decoupled it from any particular mechanism to enable the implementation of new mechanisms that specialize in certain environments.

\subsection{UCX as communication library}
In addition to being detached from any process-bootstrapping mechanism, UCX also brings benefits in performance and portability.
UCX is decoupled from either specific network hardware or programming models, which brings great compatibility and portability without compromising performance or scalability. 

Additionally, UCX provides seamless handling of Graphical Processor Unit (GPU) memory and full GPU-to-GPU direct communication, which makes it possible to accelerate applications further by GPU. 
Furthermore, Remote Direct Memory Access (RDMA) through InfiniBand and RDMA over Converged Ethernet (RoCE) is also supported by UCX, which enables some unique benefits in communication efficiency \cite{papadopoulou2017performance}. 
Firstly, UCX can make zero-copy GPU memory transfers over RDMA. 
Secondly, RDMA can significantly speed up overall communication.
For example, with UCX over RDMA on the Spark GPU cluster, the NVIDIA Rapids plugin enables a 5$\times$ reduction in time for inventory pricing queries compared with the normal Spark GPU cluster\cite{rapids}. Additionally, UCP is made to offer lightweight, portable interfaces over hardware abstraction layers and native network drivers. In order to support parallel programming models like Open-SHMEM, the UCP layer includes message layer features and protocols, including rendezvous protocols and tag matching for multi-rail networks \cite{baker2016openshmem}.

In a nutshell, UCX is a very promising communication framework that brings a lot of critical modern features to high-performance distributed computing. 
By empowering Cylon with UCX/UCC, Cylon gains wider applicable scenarios and potential performance improvement for certain working environments.

We also considered Gloo\cite{gloo} as an approach to our problem. However, UCX shows better performance results in our use case, as demonstrated in the experiment. In addition, as Gloo is built specifically for machine learning applications, UCX has a wider and more mature community in HPC.

\section{Design and Implementation}

Our work consists of mainly two parts: communication initialization and implementing the operations required by the Cylon communicator.
\subsection{Communicators overview}

The contributions described in this paper center around the Cylon communicator. The communicator is designed with the consideration that it will be implemented with multiple communication frameworks. As of now, we already have implementations with MPI, Gloo, and UCX/UCC. The communicator separates the data preparation process from the actual network-related code. Although it is effortless to achieve, not all data types are needed for each communication operator. Therefore, only selected communication operators are implemented for each data type. The availability of each data type to each communication operator is listed in table \ref{tab:ops-aval}.
\begin{table}[tbp]
\begin{center}
\begin{tabular}{|c|c|c|c|}
    \hline
    \textbf{Comm Op} & \textbf{Table} & \textbf{Column} & \textbf{Scalar} \\
    \hline
    AllGather & Available & Available & Available \\
    \hline
    Gather  & Available & Available & Available \\
    \hline
    Bcast & Available &  & \\
    \hline
    AllReduce & & Available & Available \\
    \hline
    AlltoAll & Available & &  \\
    \hline
\end{tabular}
\caption{Availability of Communication Operators on Data Types}
\label{tab:ops-aval}
\end{center}
\end{table}
\subsubsection{Serializer}
As mentioned before, different from arrays and tensors, dataframes contain heterogeneous data types and may include variable-length data types. Therefore, the communicator serializes the data into a number of buffers before passing them into the network communication layer.
There are mainly three types of data structures exchanged by communication operators: tables, columns, and scalars. As the serializer provides a set of APIs to convert tables and columns into raw data buffers, it hides the details of the data to be transferred to the communicator, making it technically straightforward to implement each communication operator for each data type.

\begin{figure}[tpb]
\begin{center}
\includegraphics[width=\linewidth]{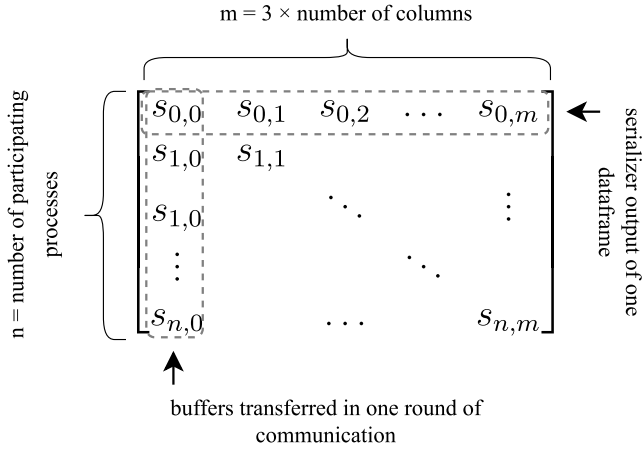}
\end{center}
\caption{Buffers to be sent produced by serializer}
\label{fig:serializer-result}
\end{figure}

Both columns and tables are serialized into a flattened data structure before communication. Cylon uses Apache Arrow to store and manage data in memory, so the serialization and deserialization processes are cohesive with the data layout in Arrow ArrayData. The serializer converts one column into three buffers: a validity bitmap, an offset array, and the data buffer. The bitmap stores the validity information, each bit in the bitmap represents the validity of one row in the column. An extra buffer stores extra bitmap information that does not align with a byte boundary. The offset array indicates the offset of each row, and it is empty for columns with fixed-width data types. The data buffer holds the data from Arrow tables. One thing to note about data buffers is that boolean data is treated with the same approach as the validity bitmap. The end result of serialization is an array of buffers with variable sizes, along with an array of each buffer’s sizes. As figure \ref{fig:serializer-result} suggests, there are $m \times n \times 3$ buffers in total, where $m$ is the number of processes involved in a communication, and $n$ is the number of columns in each process.

\subsubsection{Communication Operators}
After data is serialized into buffers, communicators begin the data-transferring process by transferring buffers' sizes. Because schemas of tables in each participating process are the same, transferring the sizes of buffers only needs fixed-sized collective operations. In \texttt{Bcast} operation, where the schema is unavailable in the receiving processes, the schema is transferred beforehand. Then, the buffers are transferred between processes with corresponding collective operations. The communication process is considered complete when all collective operations are completed.
\subsubsection{Channels and AllToAll}
Apart from the communication operators, an point to point form of communication can be achieved with the Channel API, which is used to implement the AllToAll operation. Although collective communication libraries such as MPI and UCC provide the \texttt{AllToAll} operation, we implemented it using the Channel API to give it extra flexibility. The Channel API is implemented using the \texttt{ISend} and \texttt{IRecv} functions of UCX. Other collective operations are implemented with UCC.

\subsubsection{Relationship with dataframe operators}

\begin{table}[tbp]
\begin{center}
\begin{tabular}{|c|c|}
    \hline
    \textbf{Dataframe Operator} & \textbf{Communication Operator} \\
    \hline
    Union, Difference, Join, Transpose & 
    \multirow{2}*{AllToAll}
      \\ 
    Unique, GroupBy & \\
    \hline
    Broadcast-Join & Bcast \\
    \hline
    Column-Aggregation & AllReduce \\
    \hline
    Sort & Gather, Bcast, AllToAll, AllReduce \\
    \hline
\end{tabular}
\caption{Use of communication operators in dataframe operators}
\label{tab:ops-rel}
\end{center}
\end{table}

Dataframe operators are the backbone of dataframes; a set of helpful distributed dataframe operators lays the foundation of a distributed dataframe system’s effectiveness in parallelizing massive APIs and improves developer efficiency by avoiding the redundancy among operators \cite{perera2022high}. Table \ref{tab:ops-rel} lists the dataframe operators and the communication operators they use. Our choice of dataframe operators follows a generic operator pattern, and we also choose communication operators deliberately to meet the demand of these dataframe operators. The workflow of a distributed dataframe operator usually contains the following components:
\begin{enumerate}
    \item Auxiliary local operators, such as partitioning or merging tables locally
    \item Communication operation(s)
    \item Core local operators, which are often the local version of distributed dataframe operators
\end{enumerate}
For example, the \texttt{DistributedJoin} dataframe operator mainly consists of the following process: 
\begin{enumerate}
    \item Hash target columns and split into partitioned tables
    \item Use \texttt{AllToAll} to send partitioned tables to the appropriate process
    \item Local join received tables
\end{enumerate}

\subsection{UCX/UCC Communicator}
Integrating UCX and UCC begins with implementing the Communicator and Channel. We use UCX to perform point-to-point operations in Channel and \texttt{Barrier} synchronization function and UCC to perform the collective operations. Different from other communicators such as MPI or Gloo communicator, as we use the two frameworks collaboratively with each other, we created one communicator, the UCX/UCC communicator, which embodies the functions of both of the frameworks.

When implementing the UCX/UCC communicator, we faced several challenges. The first challenge is about reusing resources. UC-Protocols (UCP) is the high-level API that UCX provides, and it can be accessed with a UCP endpoint, which contains all necessary resources for a particular network connection. Although UCC also uses the endpoints when implementing collective operations, there is no known way to reuse the endpoint we used in UCX on UCC. Therefore, we implemented the UCX/UCC communicator with two sets of UCP endpoints co-existing: one for implementing \texttt{AllToAll} with UCX and the other for the collective operations with UCC.

Another challenge is that, as of the time when we write this paper, UCC does not support Gather with variable data length in some circumstances. Therefore, we resembled the GatherV operation with the \texttt{AllGatherV} operation by allocating dummy space for receiving buffers in processes that are not gather-root. We hope that we can remove this temporary patch in the near future.

\begin{figure}[btp]
\centering
\includegraphics[width=\linewidth]{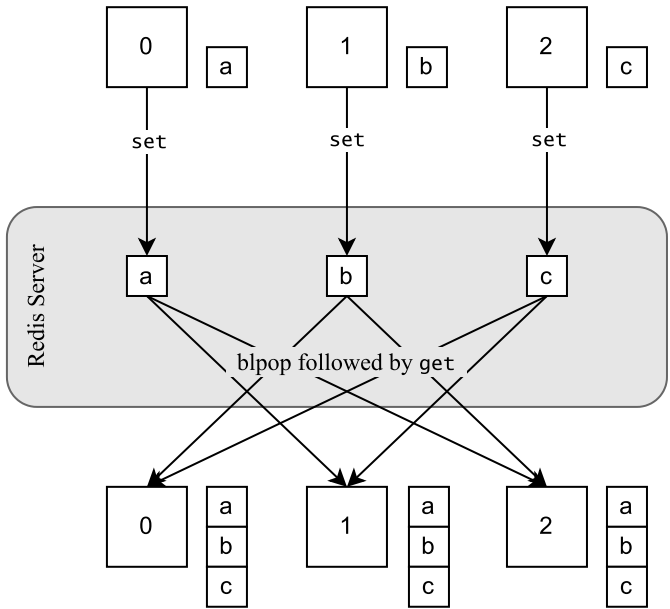}
\caption{Mimic Allgather with Redis}
\label{fig:redis-allgather}
\end{figure}

\subsection{Process bootstrapping}\label{AA}
The process initialization includes collecting communication metadata required by the communication frameworks. Metadata, including world size and rank information, is utilized in many distributed computing frameworks, including UCX and UCC. In addition, each process also needs the endpoint's address from every other process for communication to be possible. The world size, i.e., the number of processes running a specific job, is usually pre-defined and can be easily made available to each process. Therefore, we need to retrieve the rank information and the endpoint addresses for each process. 

\subsubsection{Redis}
We use an in-memory distributed key-value store, Redis, to assign a rank to communicators. The key idea is to make an atomic increment of a value served by Redis and use the value before the increment as the rank. The process of launching the Redis server is independent of the initialization process, and we only require the user to input an address of the Redis server. Therefore, the initialization process can start anywhere as long as the Redis client library can be installed.

After retrieving the world size and rank information, the UCP endpoint address of each process will be shared with every other process to enable communication using UCX and UCC. Figure \ref{fig:redis-allgather} illustrates a simplified version of this process. To retrieve the endpoint addresses, in our current approach using Redis, we map the rank of the communicator to its receiving worker address in the key-value store, and other processes retrieve the address from the store. This is to simulate the \texttt{MPI\_AllGather} operation previously used to get the address. To synchronize the processes and prevent them from retrieving endpoint addresses before they are set, each process creates an array with the length of \texttt{world\_size} after setting its address. Each process also performs a blocking pop (\texttt{blpop}) operation from other processes' arrays before retrieving (i.e., perform a \texttt{get} operation) their addresses so that the retrieving operation will be blocked until the address it tries to obtain is available. Using this process, we can prevent premature reading while avoiding the use of polling.

As mentioned previously, although UCC also utilizes the UCP endpoints to perform collective operations, we are unaware of any way to reuse the endpoints. UCC provides an API to create a UCC context, requiring a function to perform out-of-band AllGather operations. The de facto method is to use \texttt{MPI\_AllGather}. We simulated AllGather using the same idea as for UCX, with the difference that instead of passing the endpoint address information explicitly, the \texttt{ucc\_context\_create} function takes the simulated AllGather function and performs the operation without the user being aware of what data is being transferred.

We explored several other options before adopting Redis for the initialization process. For example, the Network File System (NFS) is one viable alternative. However, Redis has some advantages over NFS. First, Redis is highly available and scalable, making it ideal for distributed computing scenarios. It is also easier to perform atomic operations on Redis than on NFS since Redis is thread-safe by nature. In addition, Redis also has lower setup and maintenance costs. Another alternative we considered is ZeroMQ \cite{hintjens2011omq}, which is also fast and easy to set up. However, the tasks of distributing rank information and simulating an AllGather operation fit into the key-value model better than the publish-subscribe model.

\subsubsection{Out-of-band (OOB) contexts}
On the implementation level, the communications in the initialization process are done by using the OOB contexts, communication contexts designated to initialize UCX and UCC communication. The OOB contexts keep track of resources used in communication initialization and provide a set of APIs using which UCX and UCC communicators can gather the metadata information. In Redis OOB contexts, a Redis client instance, along with world size and rank number, are kept track of. Because multiple AllGather operations are needed to initialize UCC context, we also assign an ID to each AllGather operation so that one operation doesn’t need to wait until the previous operation to complete in all processes before it can start. This behavior is consistent with MPI, as \texttt{MPI\_Test} only tests for the completion of a local operation. Because the communication initialization process only depends on the OOB context APIs, the UCX/UCC communicator allows multiple initialization mechanisms. Initialization with MPI and Redis are currently available, but we believe that mechanisms that allow even more seamless integration with distributed computing environments can be developed in the future.

As the code snippet in Figure \ref{fig:code} shows, it only requires an address of a Redis server to initialize the communication, and the initialization process can change to use MPI easily by merely replacing the OOB context.

\begin{figure}[btp]
\vspace{0.3cm}
\begin{lstlisting}[%
 frame=single,
basicstyle=\ttfamily\footnotesize,
breakatwhitespace=false,         
breaklines=true,                 
captionpos=b,                    
keepspaces=true,                 
showspaces=false,                
showstringspaces=false,
showtabs=false,                  
tabsize=2,
language=C++
 ]
// initialize the OOB context
std::shared_ptr<net::UCCOOBContext> oob_ctx;

oob_ctx = std::make_shared<net::UCCRedisOOBContext>(
    4, "tcp://127.0.0.1:6379");

// use the line below to use MPI for OOB communication
oob_ctx = std::make_shared<cylon::net::UCCMPIOOBContext>();

/**
 * create Cylon context, which contains a 
 * communicator, with the OOB context
 */
std::shared_ptr<CylonContext> ctx;
auto ucc_config = std::make_shared<net::UCCConfig>(oob_ctx);

if (!CylonContext::InitDistributed(ucc_config, &ctx).is_ok()) {
  // unexpected
}
\end{lstlisting}
\vspace{0.3cm}
\caption{Code snippet demonstrating OOB context usage}
\label{fig:code}
\end{figure}

\section{Experiments}




Cylon UCX/UCC communication implementation was tested on a 15-node Intel\textsuperscript{\textregistered} Xeon\textsuperscript{\textregistered} Platinum 8160 cluster. Each node has a total RAM of 255GB, uses SSD for storage, and is connected via Infiniband with 40Gbps bandwidth. Each node has 48 hardware cores on 2 sockets. The software used for the experiments were: Python 3.8 \& Pandas 1.4; Cylon (Built with GCC 9.4, OpenMPI v4.1, \& Apache Arrow 5.0). UCX 1.13 was used alongside the UCC code base as of July 15, 2022. The scripts to run these experiments are available in GitHub \cite{cylon_experiments}.


Data were generated from a uniformly random distribution using NumPy, two \texttt{int64} columns. 
The first column was populated with 90\% unique keys, creating a worst-case scenario for key-based operators, such as \texttt{join} and \texttt{groupBy}. 
These NumPy data were converted to a Pandas Dataframe, which would then be converted into a Cylon Dataframes. Timings were taken only around the operators without considering the data loading times.

Two sets of experiments were carried out. We first tested the strong scaling performance of the following operator patterns discussed in our previous work \cite{perera2022high}. They were \textit{Globally reduce} (columnar-\texttt{sum}), \textit{Combine-Shuffle-Reduce} (\texttt{groupby-sum, mean, std}), and \textit{Shuffle-compute} (inner \texttt{join}). We tested the performance of dataframe operators to measure the effect of switching to UCX/UCC on our user's point of view. We didn't include operators such as \textit{Select}, \textit{Map} and \textit{Project}, because changes in communication do not affect them. We wanted to add the \texttt{sort} operation to this as well, but at the moment, the UCC library currently is missing some collective operations (e.g., \textit{Allgatherv}), and therefore, we were not able to complete it. 

The main goal of these experiments was to showcase the performance improvement achieved by integrating UCX/UCC over the existing communication implementations of Cylon. The performance of the Redis OOB communication step wasn't tested because it does not affect scaling ability, but we plan to measure the amount of overhead it contributes in the future.
    
    \subsection{Scalability of Operators (Strong Scaling)}
    
    \begin{figure*}[hptb]
    \centering
        \begin{subfigure}{.5\linewidth}
            \centering
            \includegraphics[width=\textwidth]{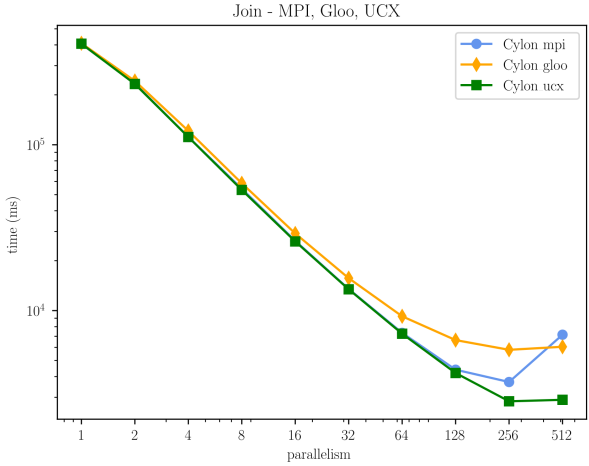}
            \caption{1B rows}
            \label{fig:join}
        \end{subfigure}%
        \begin{subfigure}{.5\linewidth}
            \centering
            \includegraphics[width=\textwidth]{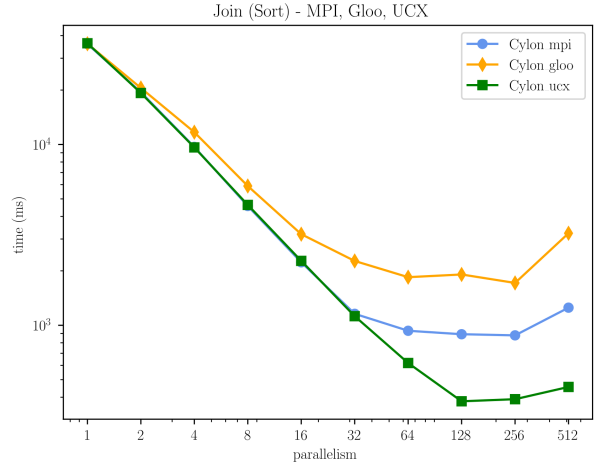}
            \caption{100M rows}
            \label{fig:join1}
        \end{subfigure}%
    \caption{Scalability of \texttt{join} (Strong Scaling) - Log-Log Plots}
    \label{fig:strong-scaling1}
    \end{figure*}
    
    \begin{figure*}[hptb]
    \centering
        \begin{subfigure}{.5\linewidth}
            \centering
            \includegraphics[width=\textwidth]{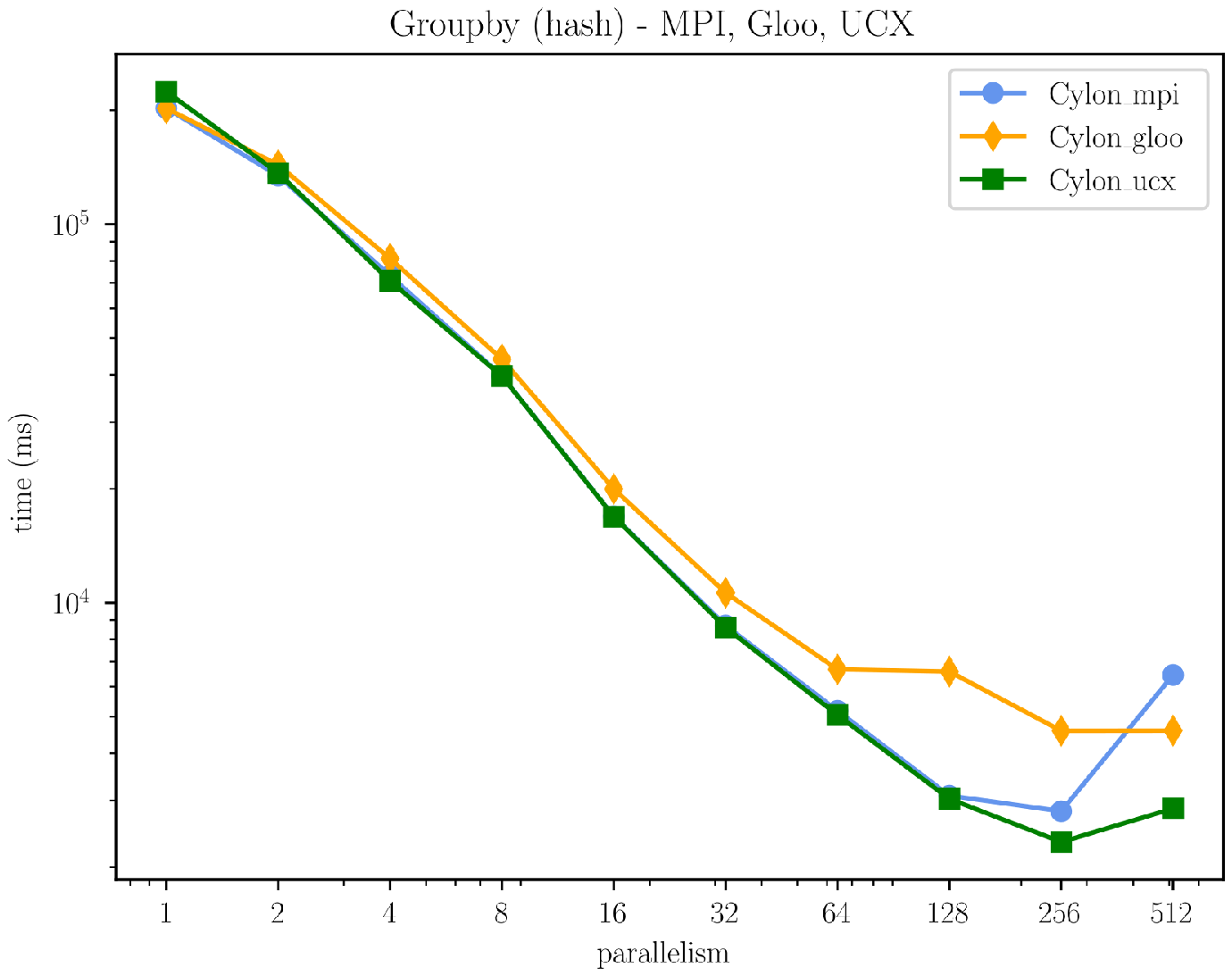}
            \caption{1B rows}
            \label{fig:gby}
        \end{subfigure}%
        \begin{subfigure}{.5\linewidth}
            \centering
            \includegraphics[width=\textwidth]{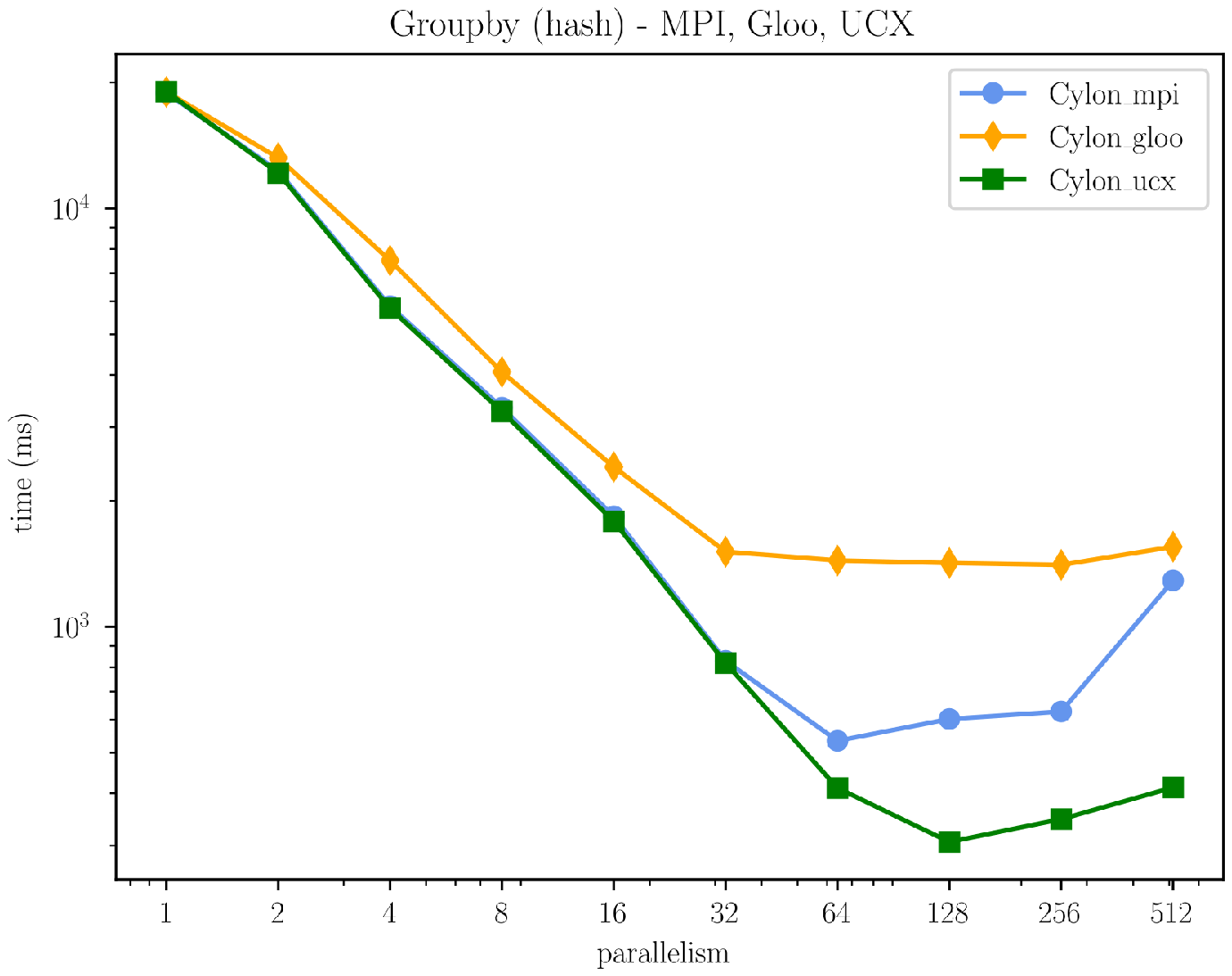}
            \caption{100M rows}
            \label{fig:gby1}
        \end{subfigure}
    \caption{Scalability of \texttt{groupby} (Strong Scaling) - Log-Log Plots}
    \label{fig:strong-scaling2}
    \end{figure*}
    
    \begin{figure*}[hptb]
    \centering
        \begin{subfigure}{.5\linewidth}
            \centering
            \includegraphics[width=\textwidth]{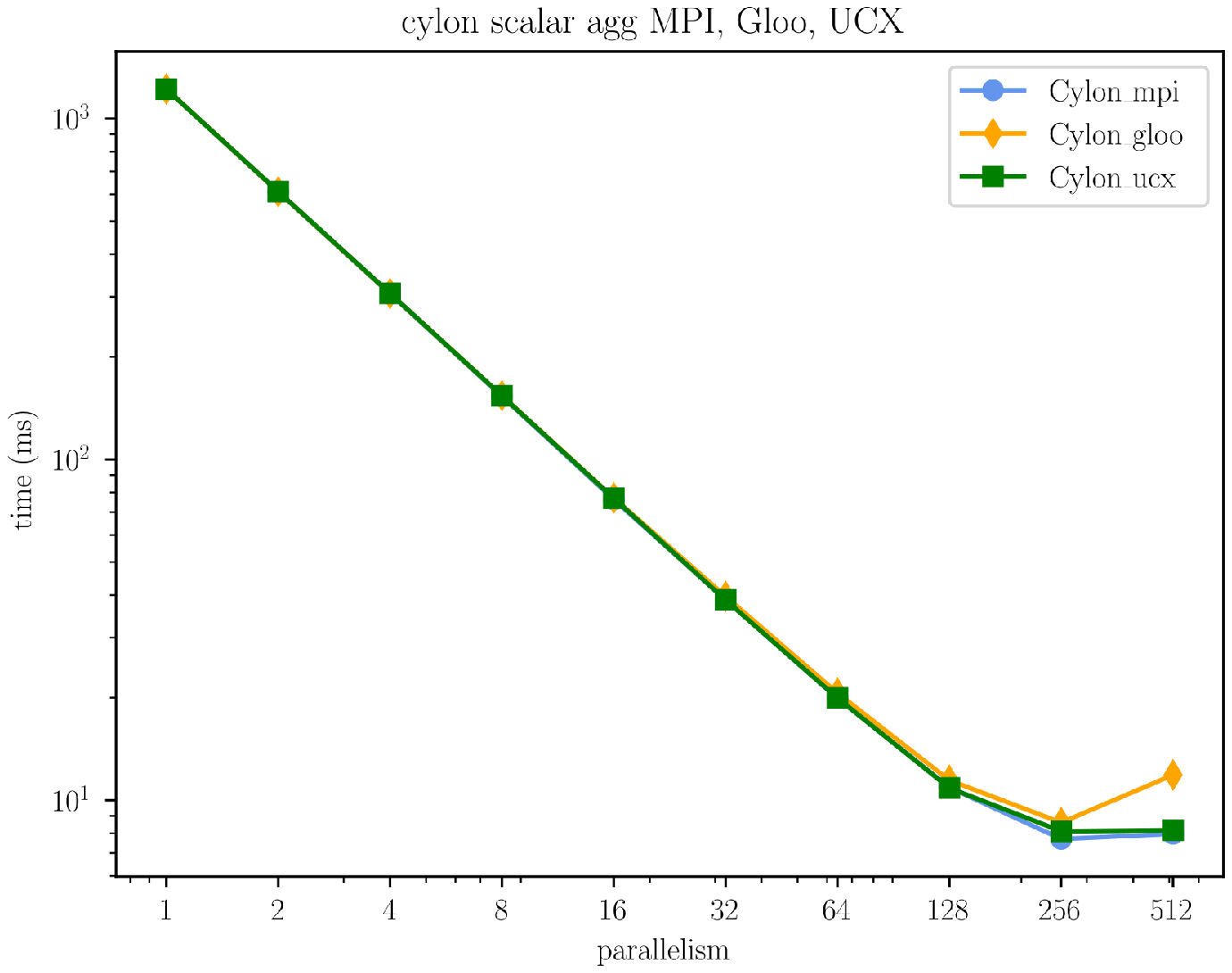}
            \caption{1B rows}
            \label{fig:agg}
        \end{subfigure}%
        \begin{subfigure}{.5\linewidth}
            \centering
            \includegraphics[width=\textwidth]{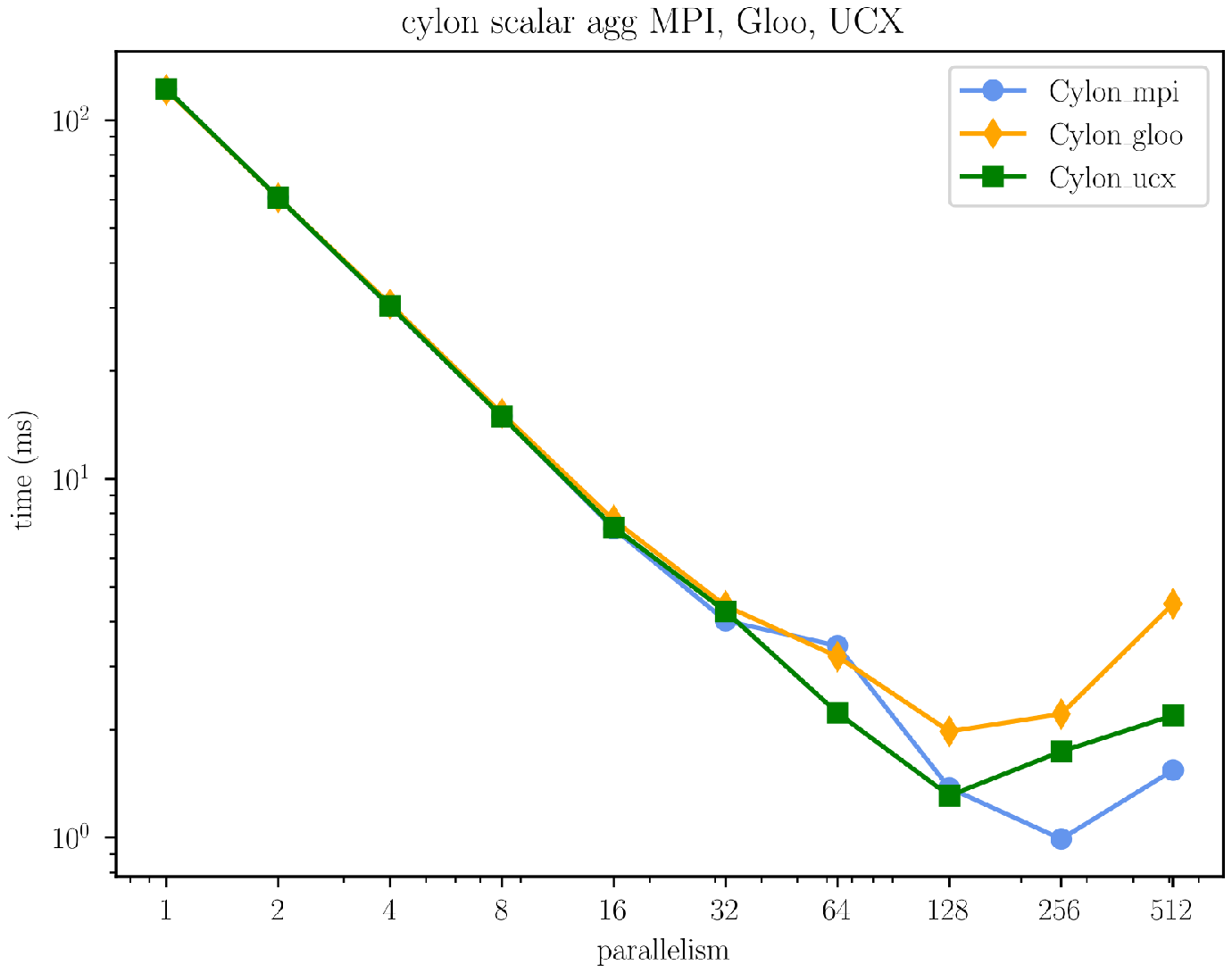}
            \caption{100M rows}
            \label{fig:agg1}
        \end{subfigure}
    \caption{Scalability of \texttt{scalar\_aggregation} (Strong Scaling) - Log-Log Plots}
    \label{fig:strong-scaling3}
    \end{figure*}
    
    Strong scaling was initially measured for $10^9$ (1 Billion) rows (roughly 16GB in size), which are shown from Figure \ref{fig:join} to Figure \ref{fig:agg}. Ideally, all these operators should follow a linear graph because the work per worker reduces as we increase parallelism. All three communication implementations seem to follow this trend. However, UCX seems to be performing better than both OpenMPI and Gloo for larger parallelism. 
    
    We further analyzed this by looking at a strong scaling plot of $10^8$ (100 Million) rows (roughly 1.6GB in size) each. This would be predominantly communication-bound, as the data size is much smaller for computation. These plots are shown from Figure \ref{fig:join1} to Figure \ref{fig:agg1}. As expected, the communication performance disparity is more pronounced in this experiment, and UCX implementation shows much better scalability than the rest. 
    
    An important point to note here is that, currently, OpenMPI integrates UCX for internal communications \cite{ompi_ucx:online}. Running our experiments using this integration would give a better idea about the communicator performance of Cylon. 

    \subsection{Weak Scaling}
        
    \begin{figure}[hptb]
        \centering
        \includegraphics[width=\linewidth]{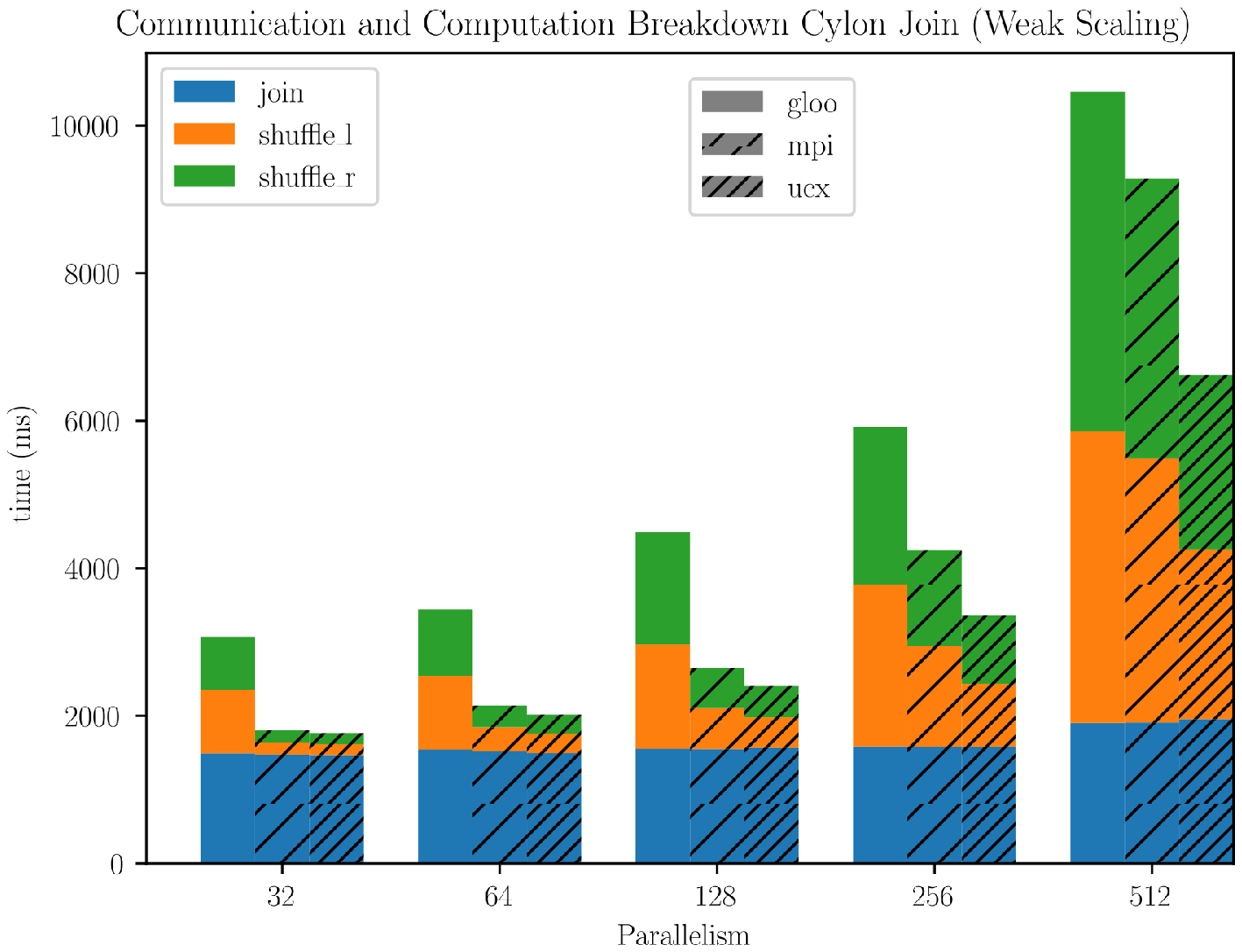}
        \caption{\texttt{join} Weak Scaling - Communication Computation Breakdown (5M rows per relation per core)}
        \label{fig:weak}
    \end{figure}
    
    To analyze the communication performance further, we carried out a weak scaling experiment for the \texttt{join} operation, with a communication and computation time breakdown. Times were measured for each hash-shuffle and the local join operation. The data set was generated with $5\times10^6$ (5 Million) rows per relation per core. Therefore, when increasing from parallelism 32 to 512, the total data size would increase from $\sim$2.5GB to $\sim$40GB.
    
    As expected, the local join time shows a nearly flat value. However, the shuffle time (i.e., the communication overhead) is increasing significantly along the parallelism axis. Gloo shows the worst performance, while the UCX communicator improves this deviation significantly. This justifies our decision to integrate UCX/UCC into Cylon. However, this experiment also shows that even though Cylon dataframe operators show decent performance, the current shuffle communication overhead might hinder the overall performance of Cylon in very large data sets and for large parallelisms. We expect to revisit the shuffle implementation with these insights as a future improvement. 
    
\section{Related Work}
{
\subsection{Data processing frameworks}

There are a lot of existent data processing frameworks designed for efficient distributed workflow. For example, the MapReduce programming model\cite{mapreduce} is one of the earliest and most famous ideas to accelerate data processing on large clusters by breaking down large data sets and processing them in parallel. Compared with MPI, the MapReduce framework can automatically parallelize user programs and provide transparent fault tolerance. MapReduce has provided a solid foundation for later successors, among which Spark\cite{zaharia2010spark} gradually dominates. As an in-memory processing framework, Spark can provide potentially a 10$\times$ speedup compared with MapReduce by avoiding writing back to disks during processing. Thanks to its high performance and low-latency data sharing, Spark is much more efficient with multi-pass applications like streaming processing, SQL, machine learning, etc. Although Spark benefits from its wide application scenes, it is not necessarily the best solution for some specific applications. Spark leverages micro batches to emulate streaming, which requires a careful trade-off between throughput and latency. Apache Flink\cite{Carbone2015ApacheFS} is yet another data processing framework that supports native streaming so that higher throughput and consistency are guaranteed. Additionally, Apache Flink has more support for iterative processing like machine learning by native loop operators which is absent in Spark.



\subsection{Dataframes}
    Due to the rapid growth of dataset in a variety of field of data engineering, the use of heterogeneous data on single or multiple nodes push hard to extend its current limit. \textit{Dataframe(DF)} and \textit{Distributed Dataframe(DDF)} are the core unit of handling large-scale data engineering applications. Compared with the relational database table, DFs consist of homogeneous or heterogeneous data \cite{abiteboul1995foundations}. DF adaptation with heterogeneity distinguishes it from multidimensional arrays or tensors.
    DDFs solve the problem by acting as a unit placed in thousands of nodes concurrently and provides a common infrastructure on the cloud to handle billions of heterogeneous data for different ML or HPC applications and save significant development time (nearly 60\%) \cite{acds2021}. In addition, Modin offers a pandas-like API that employs Ray or Dask to create a framework for high-performance distributed execution. On a computer with four physical cores, it offers speedups of up to four times. \cite{petersohn2021flexible}.

    Partitioned datasets with well-separated blocks along an index can be effectively computed using the dask dataframe. Users gain from Dask dataframe's straightforward access to larger-than-memory datasets and concurrent computing in cases where Pandas does release the Global Interpreter Lock (GIL) \cite{pythonx2013Simplified}. The higher-level collections dask dataframe for general computing shows the adaptability of the dask graph specification to encode complex parallel algorithms and the capacity of the dask schedulers to execute those graphs on a multi-core computer intelligently. Though it doesn't neatly fit into a single high-level abstraction like arrays or dataframes and is instead merely a collection of related Python functions with data dependencies, dask graph allows for parallel execution to go beyond \texttt{ndarrays} and dataframes \cite{rocklin2015dask}. The dask runtime cannot yet be added to a distributed environment with a protocol-independent framework.

    CuDF is a Python GPU DataFrame framework for reloading, joining, aggregating, filtering, and manipulating tabular data using a DataFrame style API. Where necessary, Dask-cuDF upgrades Dask to enable cuDF GPU DataFrames to process its DataFrame partitions rather than Pandas DataFrames. Dask-cuDF is advantageous if the workflow is spread across numerous GPUs, has more data than can fit in memory on a single GPU, or has to analyze data scattered across multiple files at once \cite{rapidsWelcomeCuDFx2019s}.

    
    Compared with the frameworks mentioned above, Cylon empowered by UCX \& UCC with distributed dataframe is designed for general high-performance computing environments.
}

\section{Future Works}
Although we constructed a new communication initialization mechanism using Redis for environments where MPI is unavailable, we anticipate that a significant portion of Cylon’s use cases will still be in environments with MPI. Therefore, it would be a minor inconvenience for some users if Redis becomes a strict dependency. In addition, as MPI’s bootstrapping mechanism is more holistic, we prefer to use MPI for bootstrapping when possible. Therefore, we intend to enable a runtime detection of MPI’s presence in the environment and use it as the bootstrapping mechanism when feasible. As mentioned previously, the \texttt{Gather} communication operator with variable data length is not supported as of the time this paper is composed, and we made a workaround by emulating \texttt{Gather} with the \texttt{AllGather} operation. We intend to remove this workaround when GatherV is available in UCC. Moreover, we also want to find a way to reuse the UCP endpoints to conserve resources.

Additionally, as mentioned before and depicted in Figure \ref{fig:weak}, the \texttt{join} operation shows a weak scaling pattern in performance. Currently, the \texttt{shuffle} operation is executed with point-to-point communication operations, instead of the \texttt{AllToAll} operation from communication libraries. We made this decision because the Tables become non-contiguous in memory after \texttt{hash partition}, and using collective operations requires an extra copy to place the table data in the contiguous memory. As the implementation with point-to-point communications is more costly in performance than with collective operations, we intend to refactor the \texttt{shuffle} operation. This will require us to make \texttt{hash partition} in-place and use \texttt{AllToAll} operation to transfer the partitioned data to their corresponding workers.

\section{Conclusion}
In summary, this paper presents the integration of distributed-memory parallel dataframe, Cylon, and communication framework UCX. To our knowledge, this is the first attempt of directly integrating UCX in distributed dataframes. We present the challenge we faced with the dependency on MPI and introduce a workaround to mitigate this issue and make integrating with Cylon easier for some distributed computing environments. We presented an overview of Cylon to provide context, and we showed the effort we made to support this integration. We envision that, with the direct integration of UCX, Cylon can achieve more promising performance results and also expand its surface area to cover more environments and use cases. Although this work builds upon Cylon, we believe that integrating directly with UCX can be a general solution to narrow the gap between MPI-based and MPI-incompatible applications.

\section*{Acknowledgment}
We gratefully acknowledge the support of NSF grants 2210266 (CINES) and 1918626 (GPCE).

\bibliographystyle{IEEEtran}   
\bibliography{ref.bib}
\end{document}